\documentclass{Interspeech}

\interspeechcameraready

\title{Model as Loss: A Self-Consistent Training Paradigm}

\author[affiliation={1}]{Saisamarth Rajesh}{Phaye}
\author[affiliation={1}]{Milos}{Cernak}
\author[affiliation={1}]{Andrew}{Harper}

\affiliation{}{Audio Machine Learning}{Logitech}

\email{(sphaye, mcernak, aharper)@logitech.com}

\usepackage{comment}
\usepackage{amsmath}
\usepackage{arydshln}
\usepackage{graphbox}
\usepackage{balance}
\begin{document}

\maketitle

\begin{abstract}
    
    Conventional methods for speech enhancement rely on handcrafted loss functions (e.g., time or frequency domain losses) or deep feature losses (e.g., using WavLM or wav2vec), which often fail to capture subtle signal properties essential for optimal performance. To address this, we propose Model as Loss, a novel training paradigm that utilizes the encoder from the same model as a loss function to guide the training. 
    
    The Model as Loss paradigm leverages the encoder's task-specific feature space, optimizing the decoder to produce output consistent with perceptual and task-relevant characteristics of the clean signal. By using the encoder's learned features as a loss function, this framework enforces self-consistency between the clean reference speech and the enhanced model output. Our approach outperforms pre-trained deep feature losses on standard speech enhancement benchmarks, offering better perceptual quality and robust generalization to both in-domain and out-of-domain datasets.
\end{abstract}

\section{Introduction}

Speech enhancement has long been a challenging problem, with applications in telecommunication, hearing aids, and robust automatic speech recognition (ASR) \cite{benesty2006speech, reddy2020interspeech, babaev2024finally, dubey2023icassp2023deepnoise, deepfilternet2}. A critical component in training enhancement models is the choice of loss function, which directly influences the quality and generalization of enhanced output \cite{deepfeaturelosses, braun2021consolidated}. Conventional loss functions, such as time or spectrogram-domain losses \cite{steinmetz2020auraloss}, often do not fully capture the complex relationships between noisy and clean speech.
For instance, spectrogram loss treats all frequency bins equally, which can result in overemphasis of less perceptually relevant regions while underweighting crucial frequencies important for speech intelligibility \cite{pmsqe}. Although there are perceptually sensitive losses such as mel-spectral losses \cite{steinmetz2020auraloss} or PMSQE \cite{pmsqe}, they overly compress the signal, limiting the preservation of fine-grained details crucial for maintaining speech intelligibility, achieving suboptimal performance \cite{kolbaek2020loss}.

Pre-trained deep feature losses \cite{deepfeaturelosses}, such as those derived from WavLM \cite{wavlm} or Wav2Vec \cite{schneider2019wav2vec}, have gained popularity due to their ability to incorporate perceptual and contextual information into training objectives. Recent work by Babaev \textit{et al.} \cite{babaev2024finally} shows that WavLM's intermediate convolutional features have a high correlation to speech enhancement, as compared to its transformer layers. WavLM has also been shown to be superior to Wav2Vec 2.0 \cite{wav2vec2} when used as a loss function. However, these methods are often optimized for tasks such as ASR or phoneme recognition, which may not align with speech enhancement objective. Such losses might prioritize linguistic content while ignoring residual noise components critical to the enhancement task. Moreover, these pre-trained neural networks, when used as loss functions, can suffer from limited sensitivity to noise, as the extracted features may focus on abstract representations rather than task-specific properties \cite{maiti2022speechlmscoreevaluatingspeechgeneration}.

\begin{figure*}[t]
    \centering
    \makebox[\textwidth][c]{
    \begin{tabular}{@{}lcccccc@{}}
        \toprule
        
        & Baseline & 
        Baseline$_{\text{10epochs}}$ &
        Baseline$_{\text{wavlm}}$ &
        \textbf{Ours$_{\text{mal}-\text{frozen-fe}}$} &
        \textbf{Ours$_{\text{mal}-\text{frozen}}$} &
        \textbf{Ours$_{\text{mal}-\text{dynamic}}$} 
        \\ 
        \midrule
        Noisy Spectrogram & 
        \includegraphics[align=c,width=0.12\linewidth]{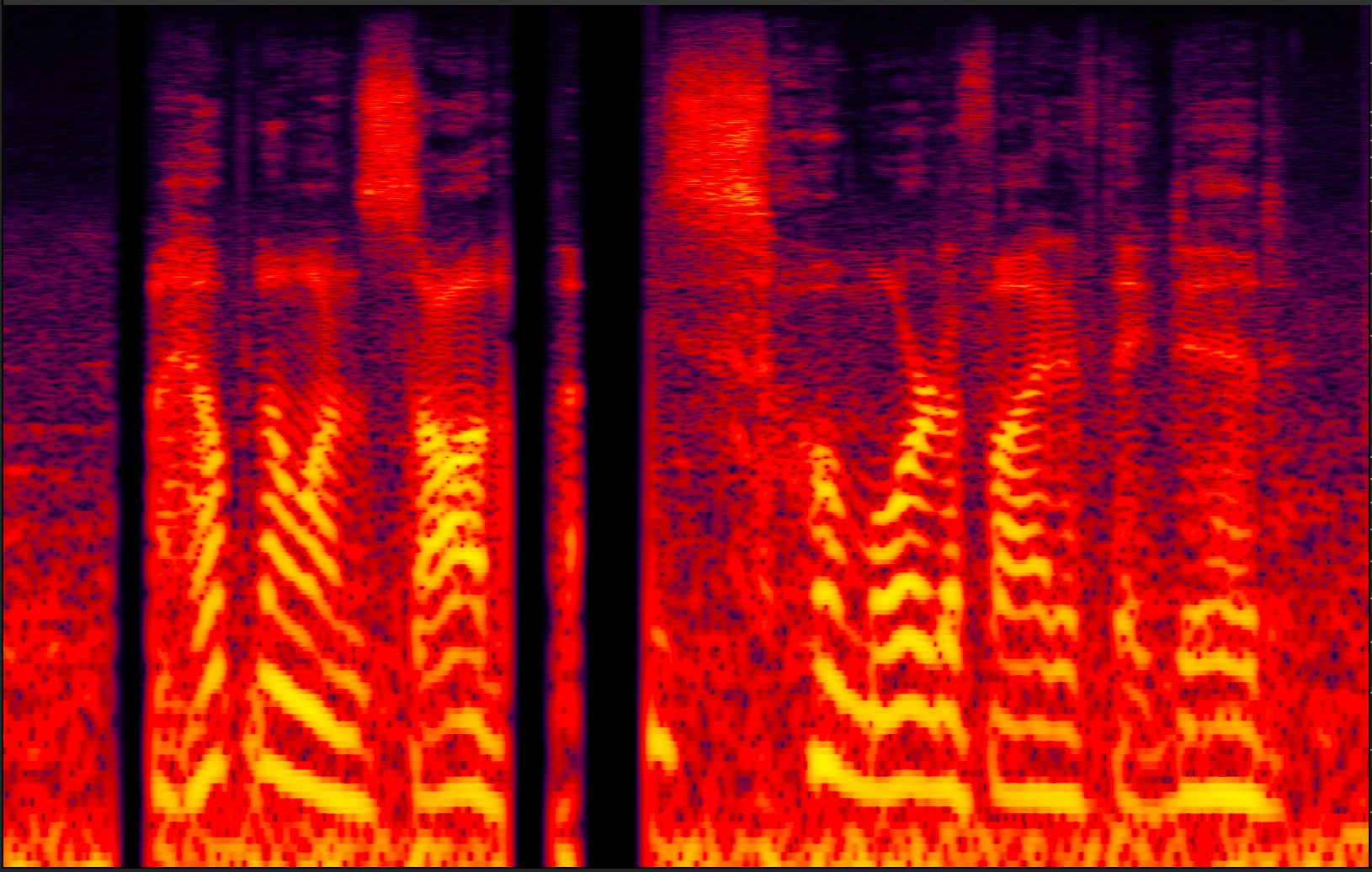} & 
        \includegraphics[align=c,width=0.12\linewidth]{images/noisy-min.png} & 
        \includegraphics[align=c,width=0.12\linewidth]{images/noisy-min.png} & 
        \includegraphics[align=c,width=0.12\linewidth]{images/noisy-min.png} & 
        \includegraphics[align=c,width=0.12\linewidth]{images/noisy-min.png} & 
        \includegraphics[align=c,width=0.12\linewidth]{images/noisy-min.png} \\ 
        
        \midrule
        
        1 Enhancement & 
        \includegraphics[align=c,width=0.12\linewidth]{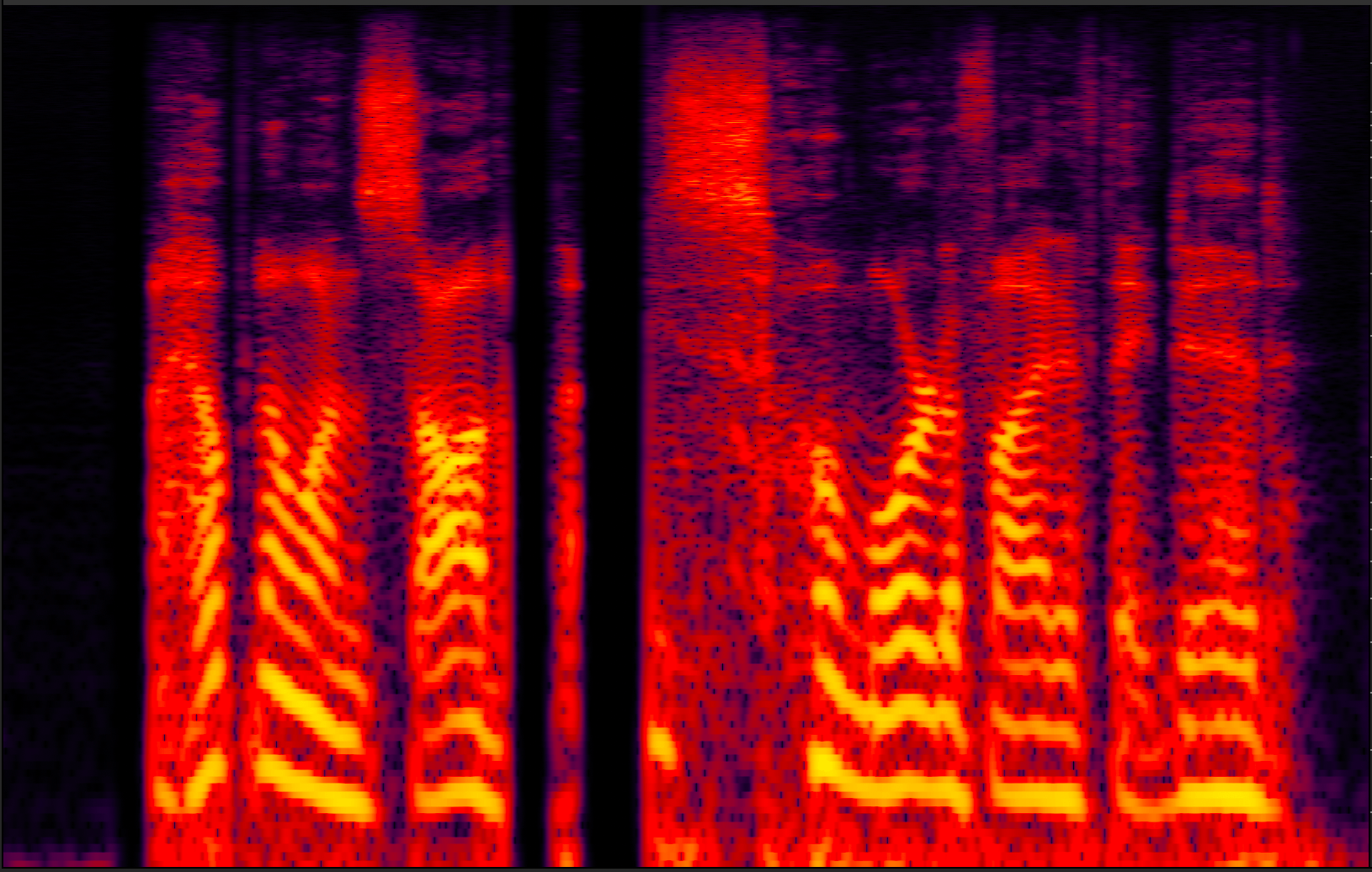} & 
        \includegraphics[align=c,width=0.12\linewidth]{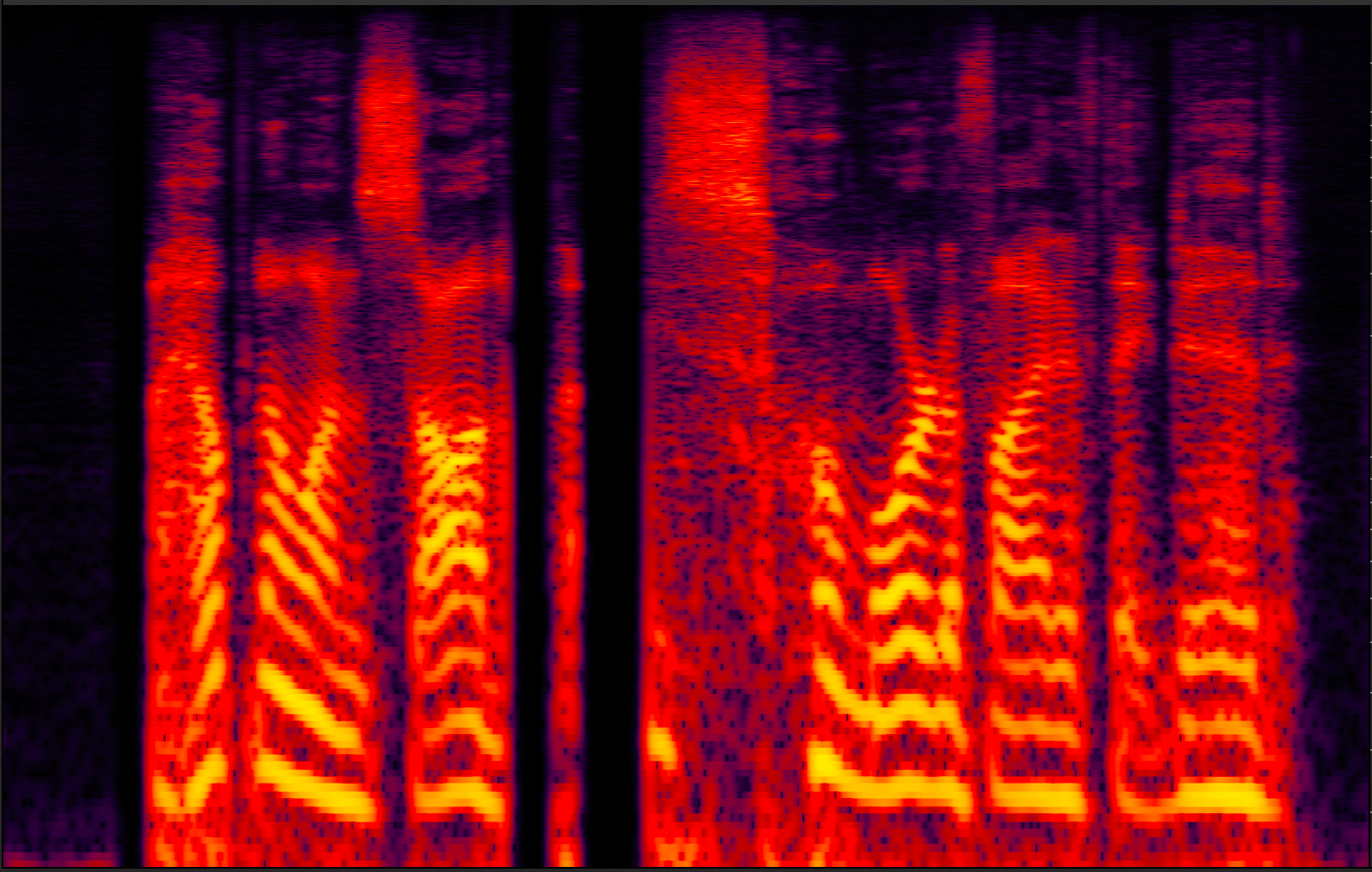} & 
        \includegraphics[align=c,width=0.12\linewidth]{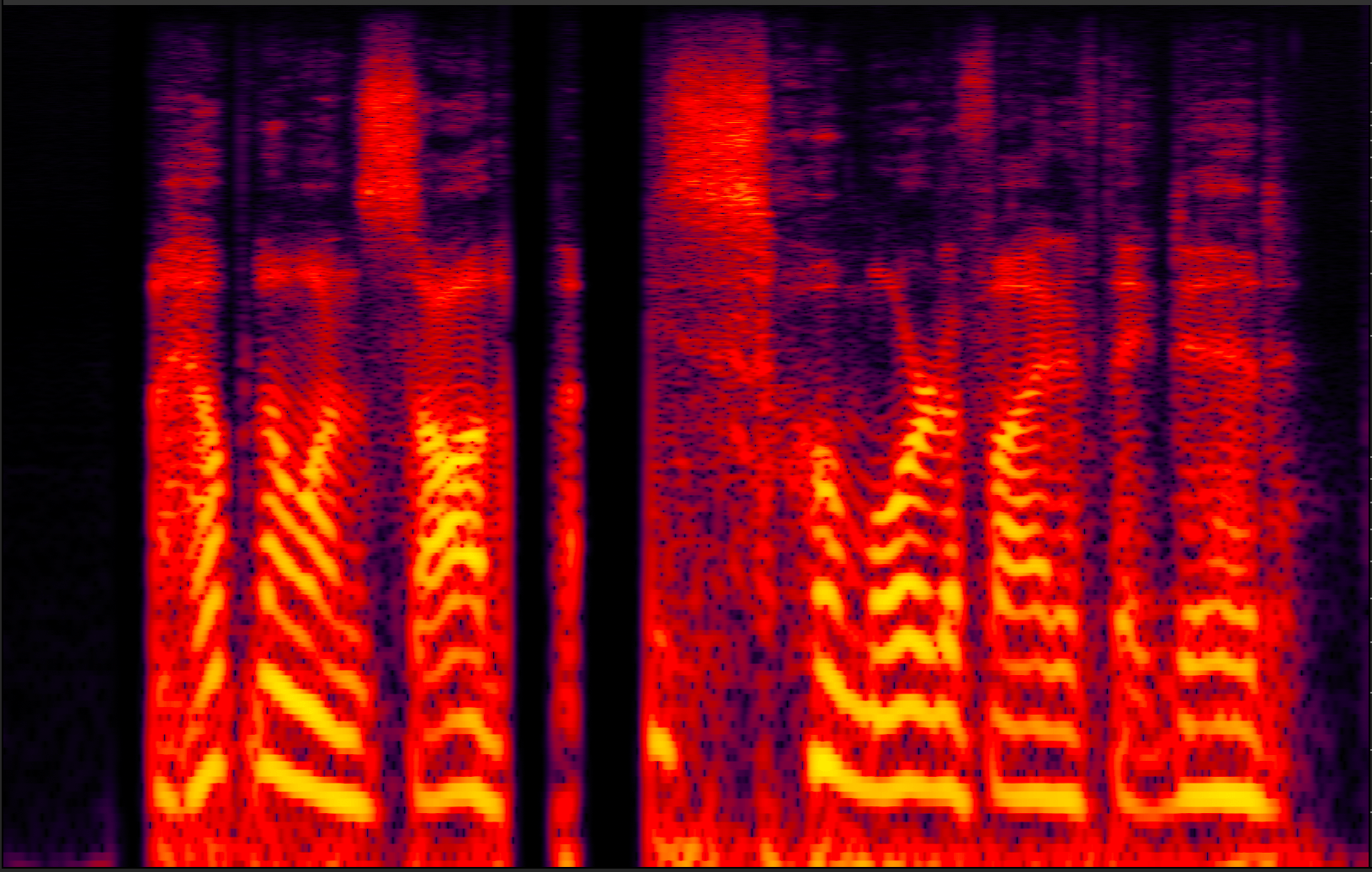} & 
        \includegraphics[align=c,width=0.12\linewidth]{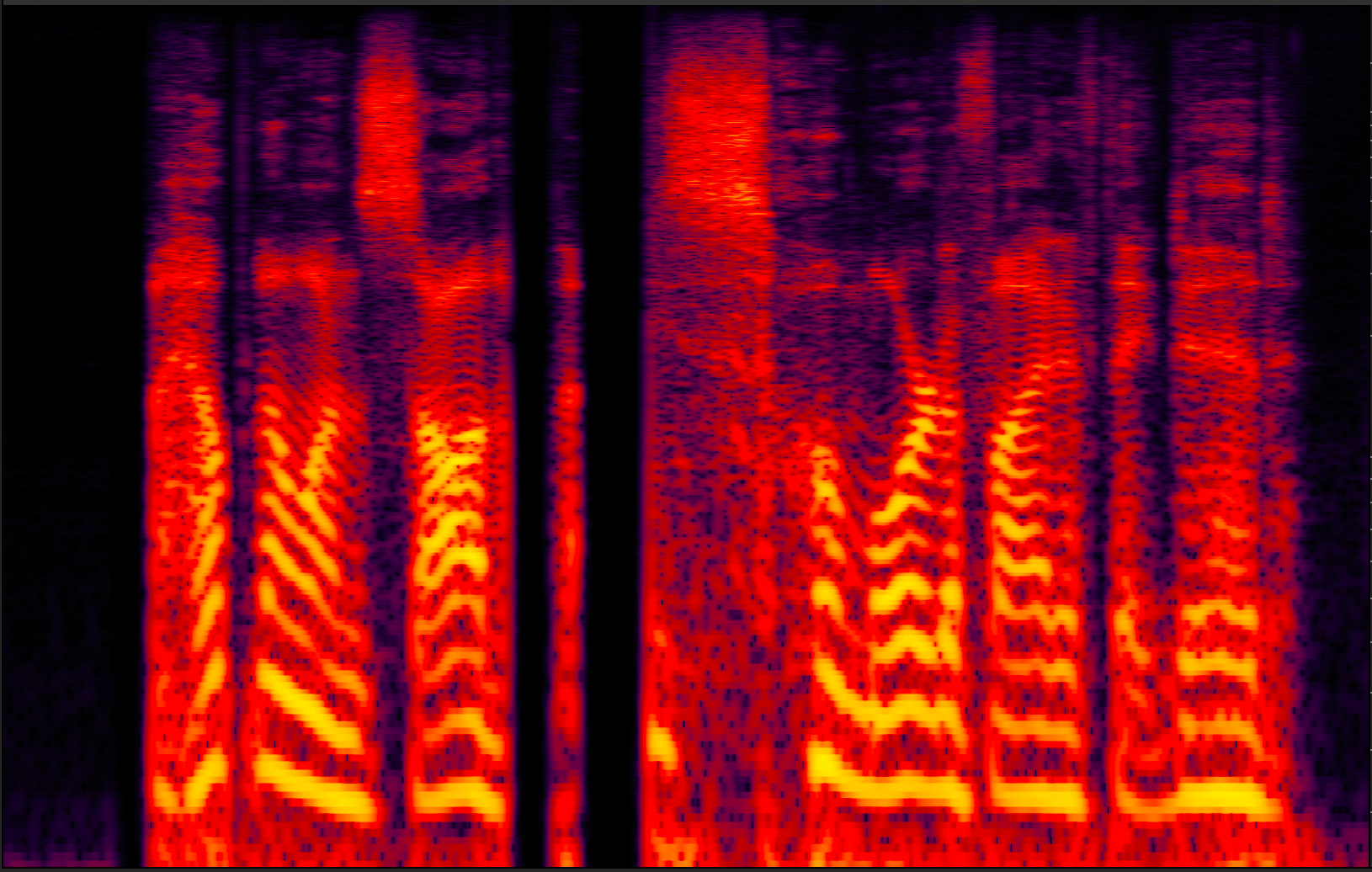} & 
        \includegraphics[align=c,width=0.12\linewidth]{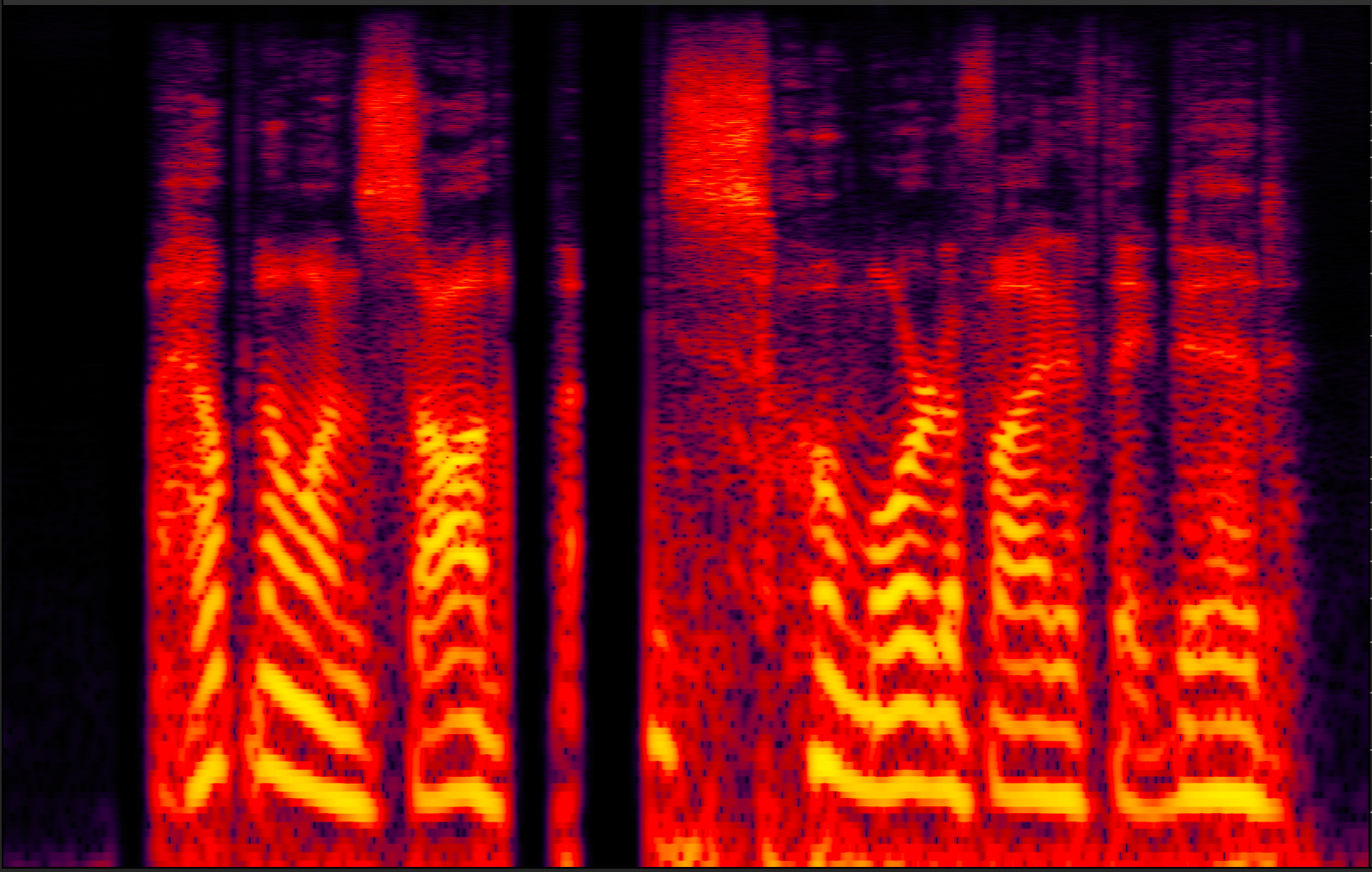} & 
        \includegraphics[align=c,width=0.12\linewidth]{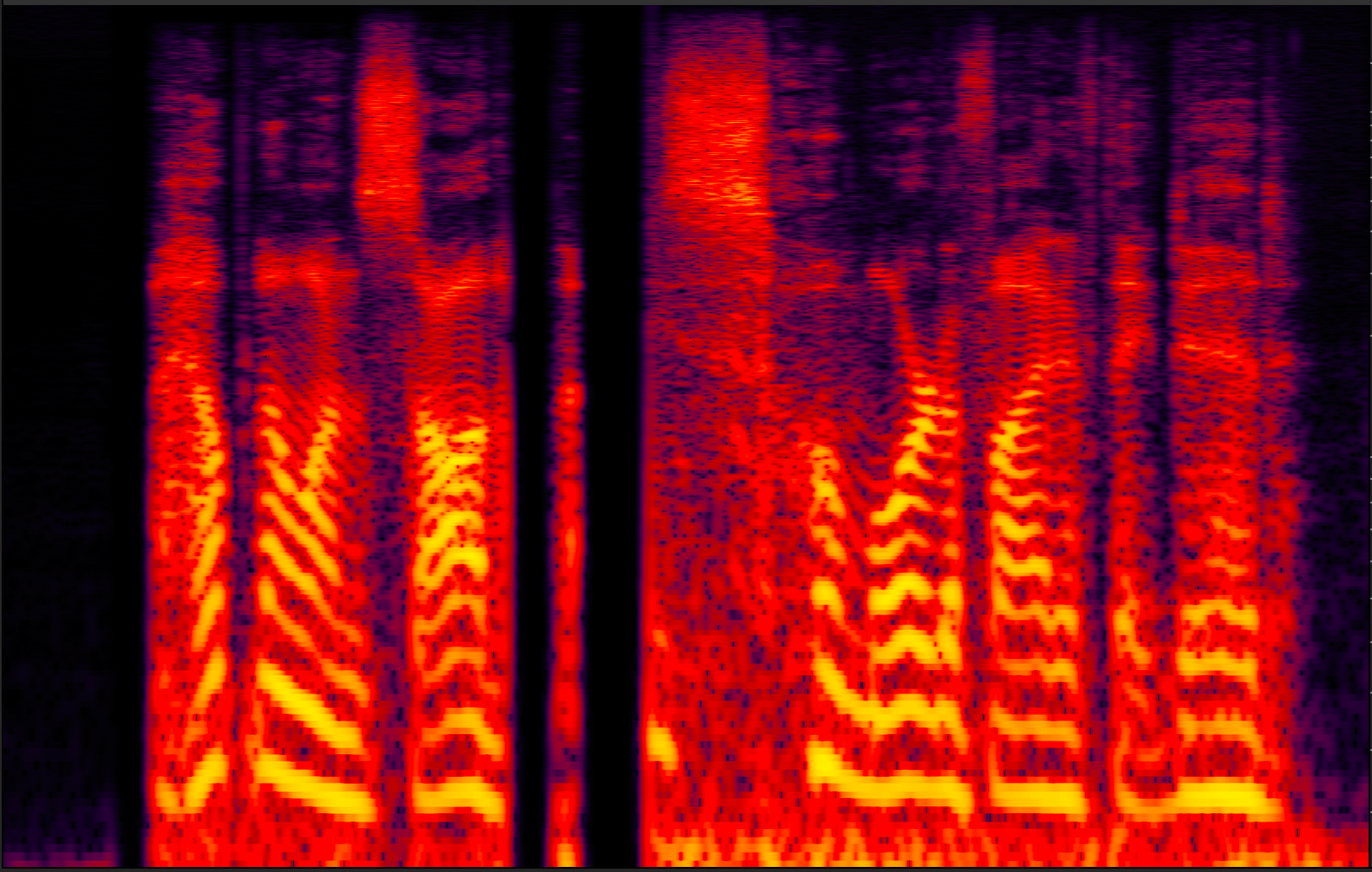} \\ 
        
        10 Enhancements & 
        \includegraphics[align=c,width=0.12\linewidth]{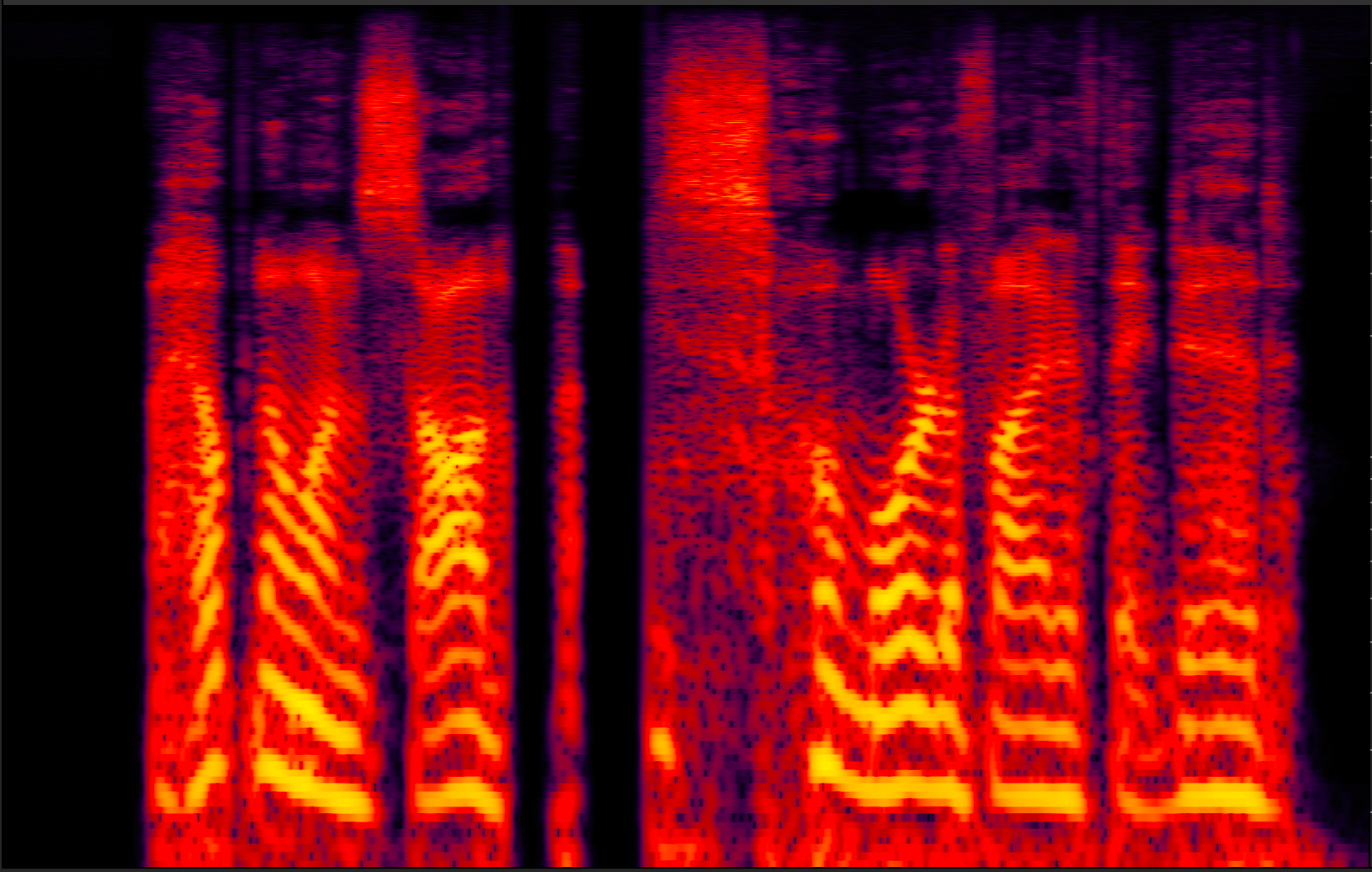} & 
        \includegraphics[align=c,width=0.12\linewidth]{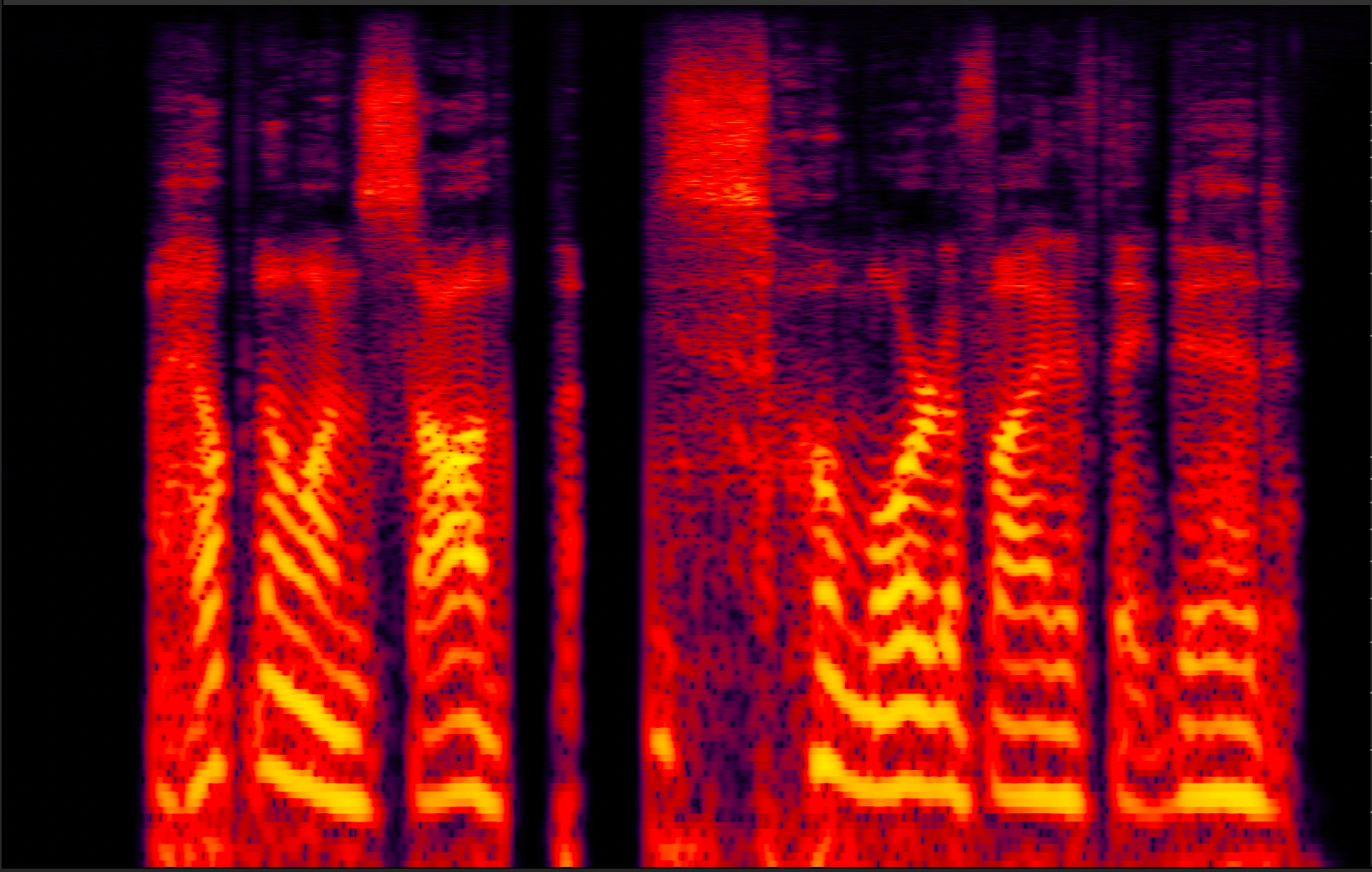} & 
        \includegraphics[align=c,width=0.12\linewidth]{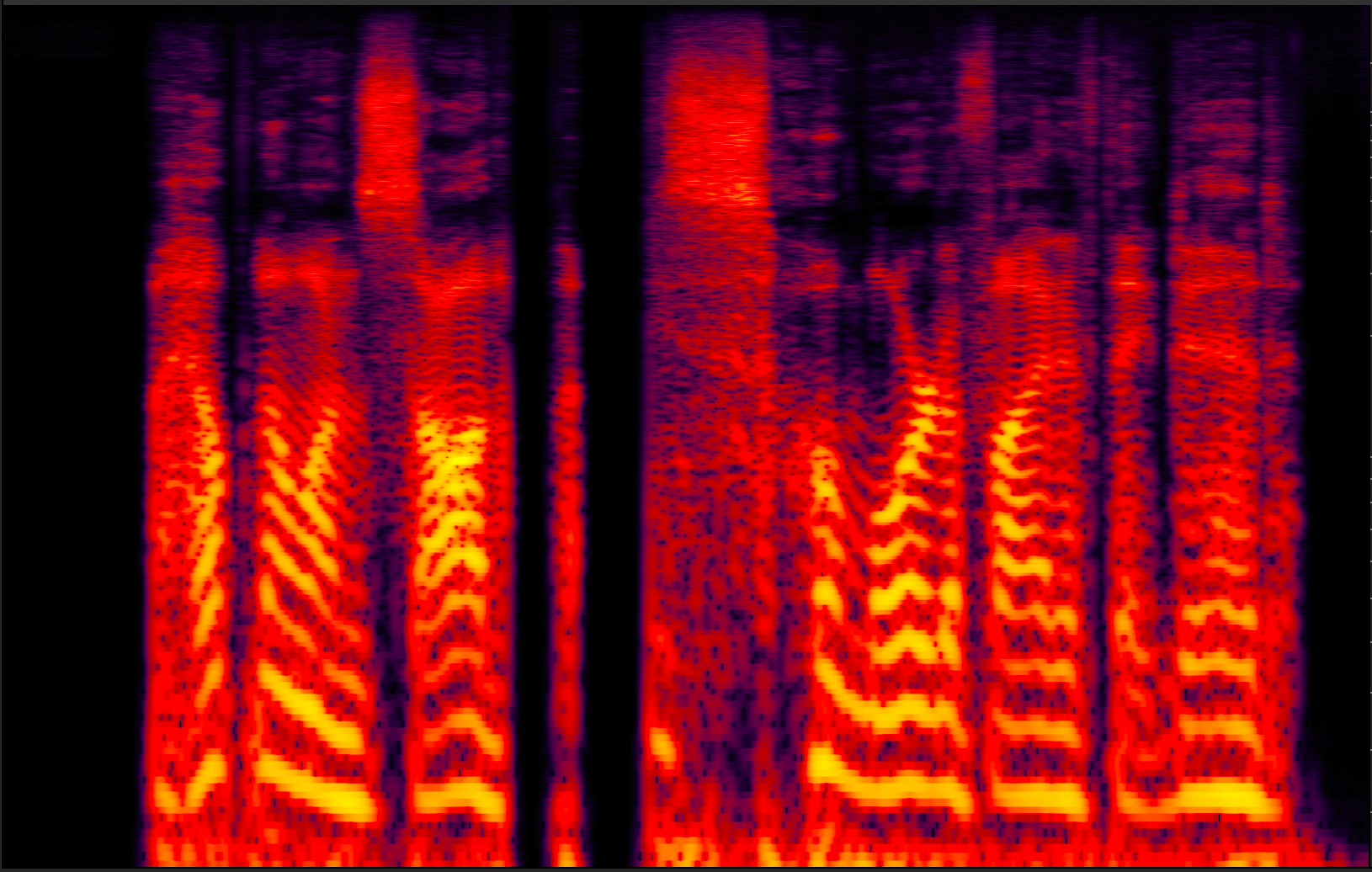} & 
        \includegraphics[align=c,width=0.12\linewidth]{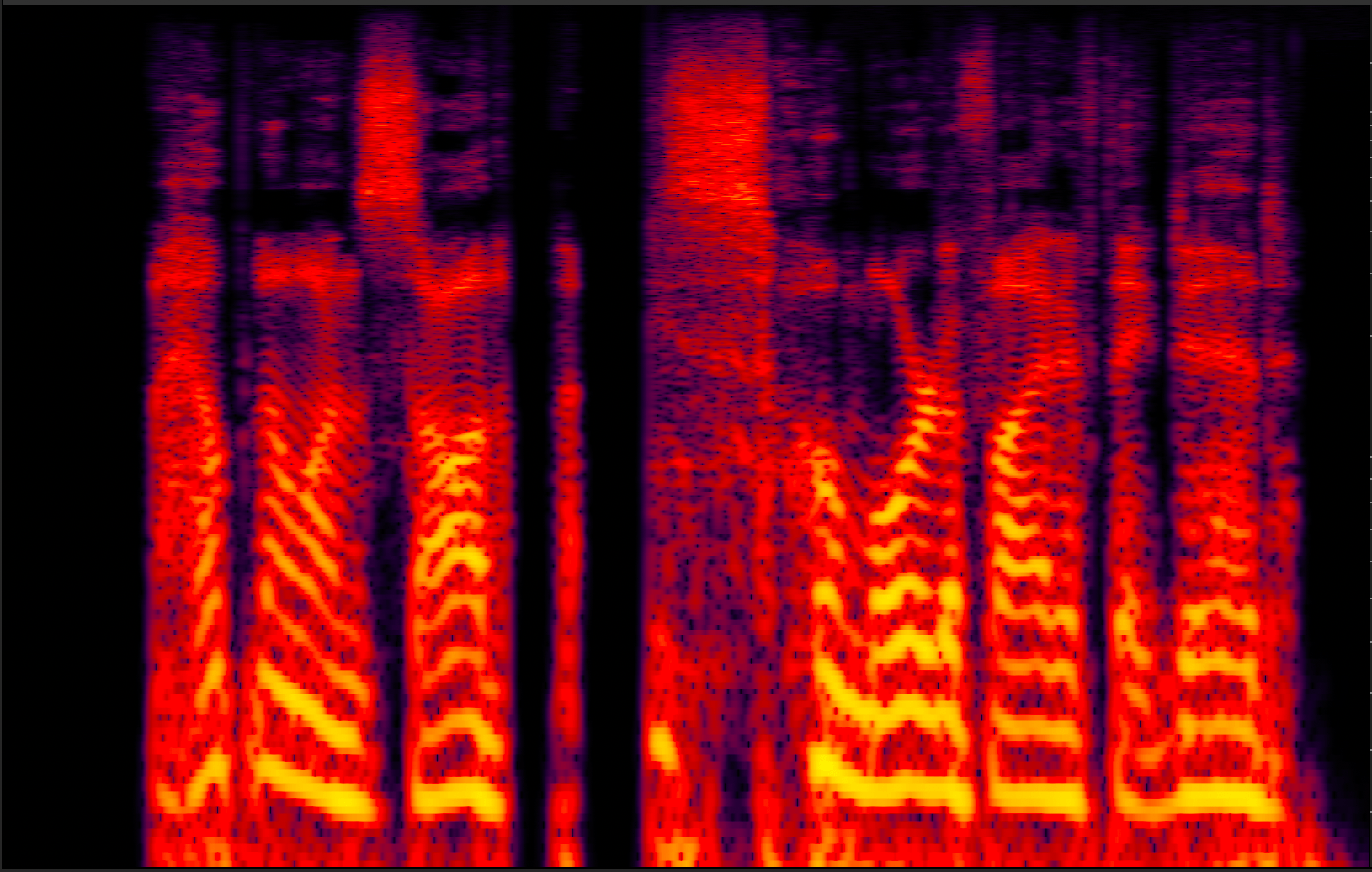} & 
        \includegraphics[align=c,width=0.12\linewidth]{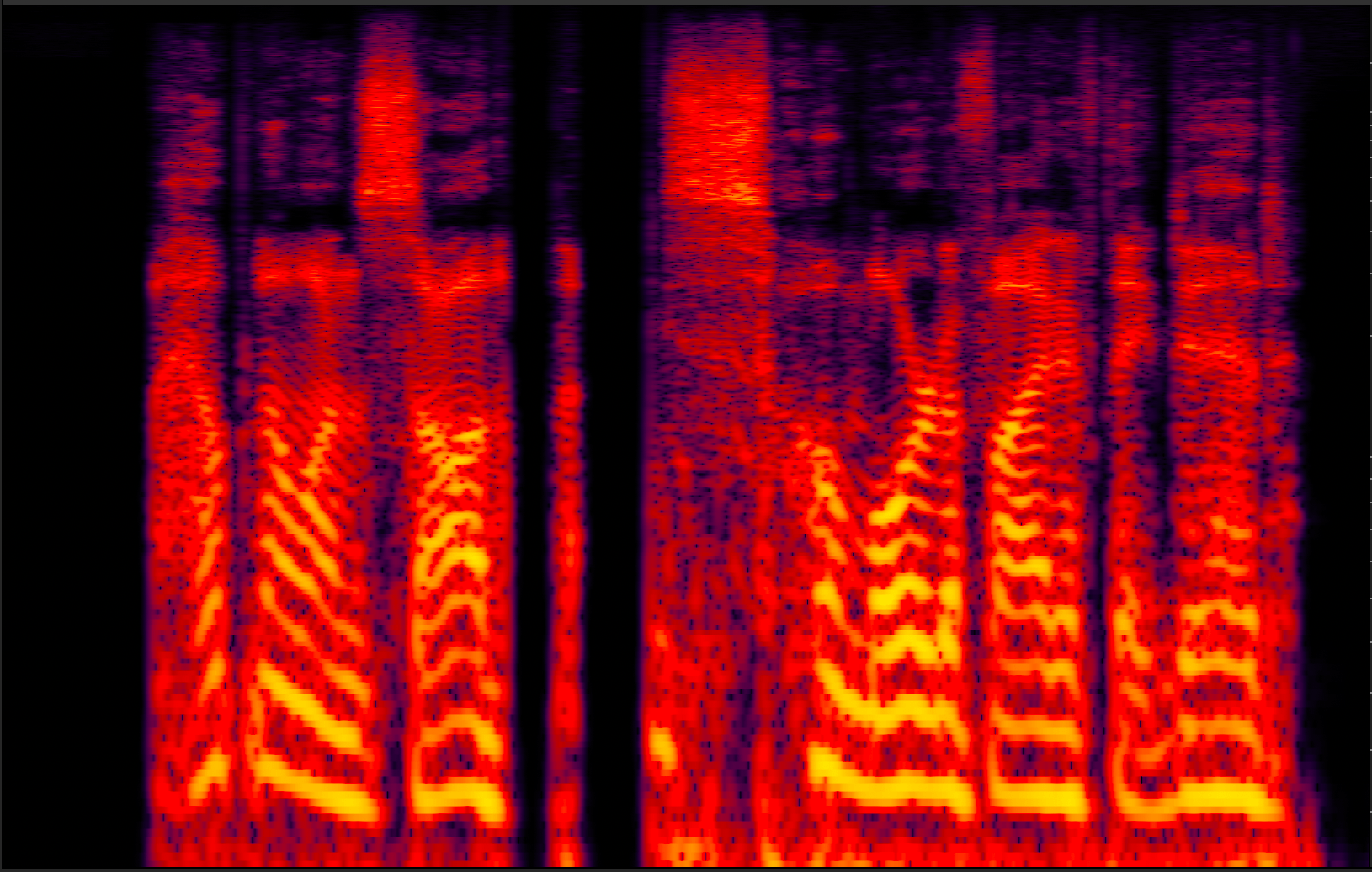} & 
        \includegraphics[align=c,width=0.12\linewidth]{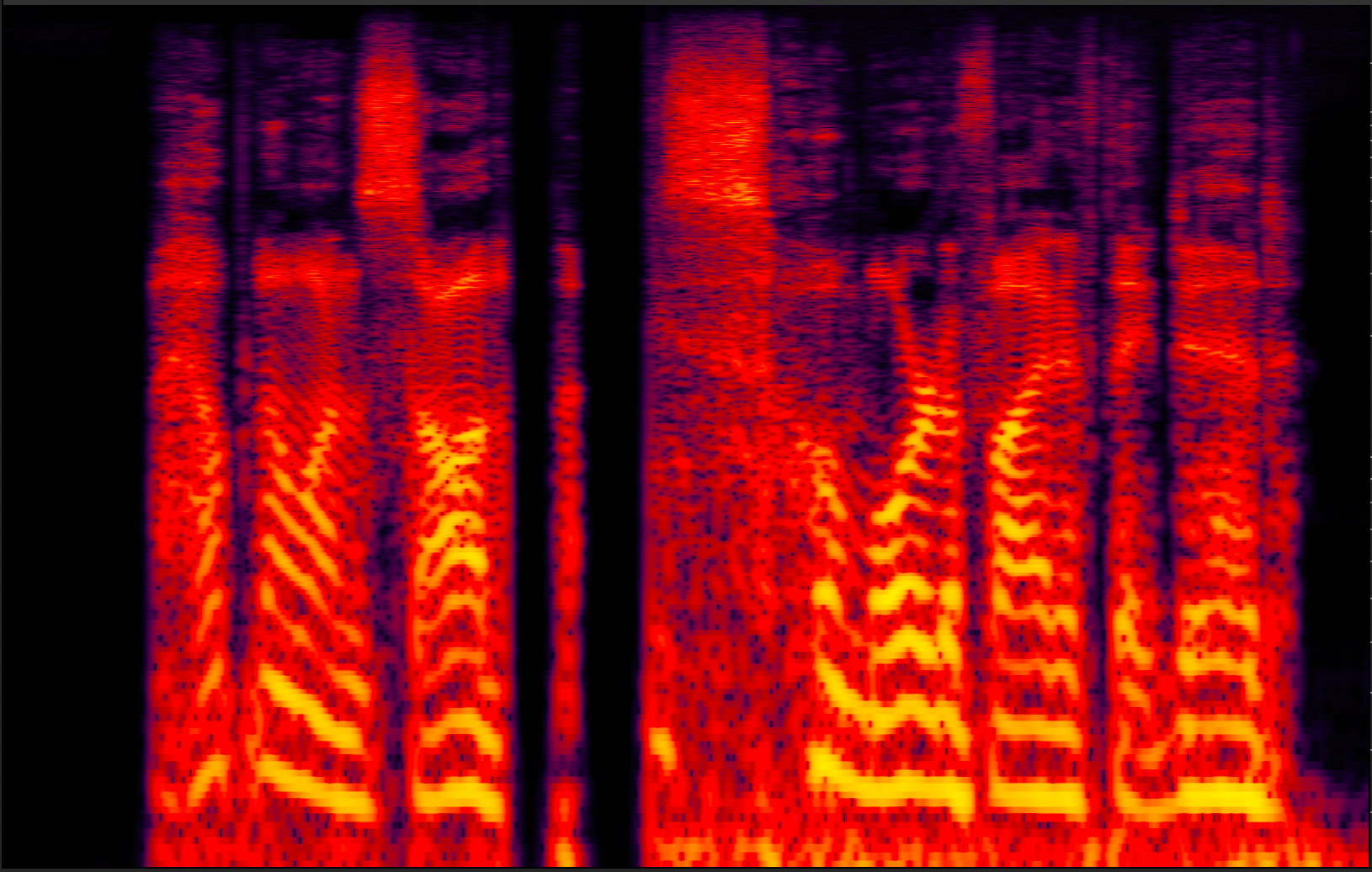} \\ 
        
        
        100 Enhancements & 
        \includegraphics[align=c,width=0.12\linewidth]{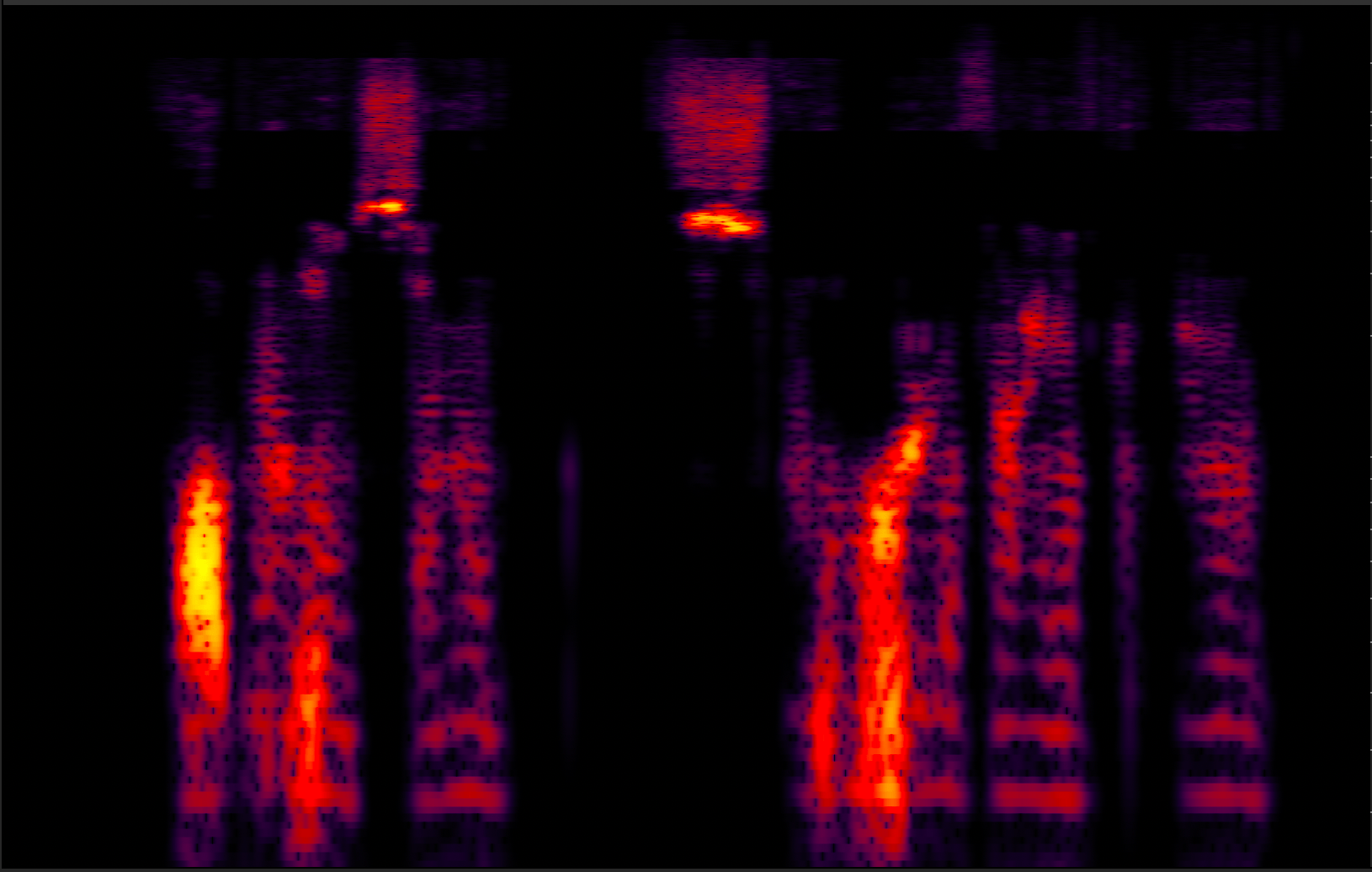} & 
        \includegraphics[align=c,width=0.12\linewidth]{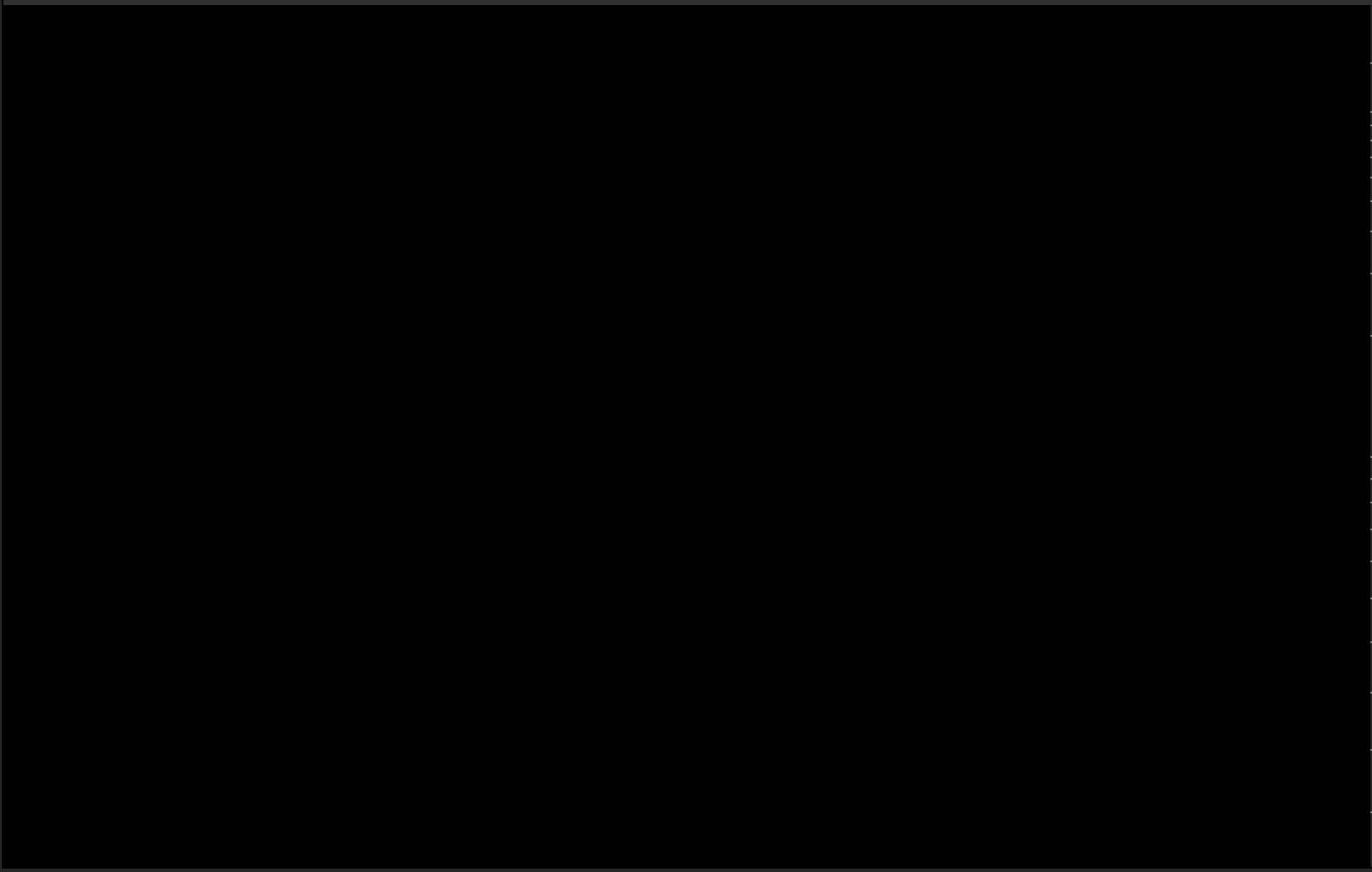} & 
        \includegraphics[align=c,width=0.12\linewidth]{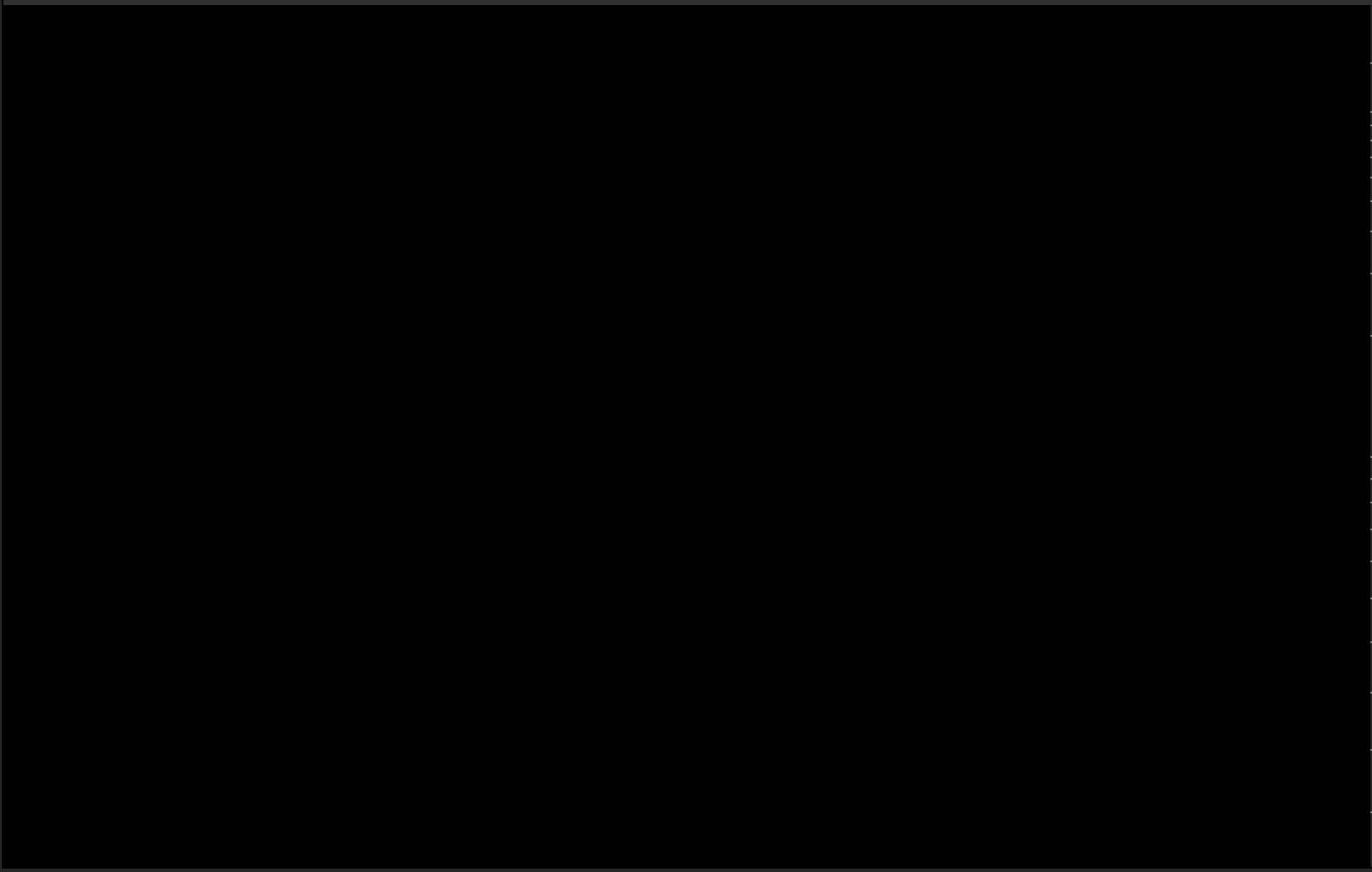} & 
        \includegraphics[align=c,width=0.12\linewidth]{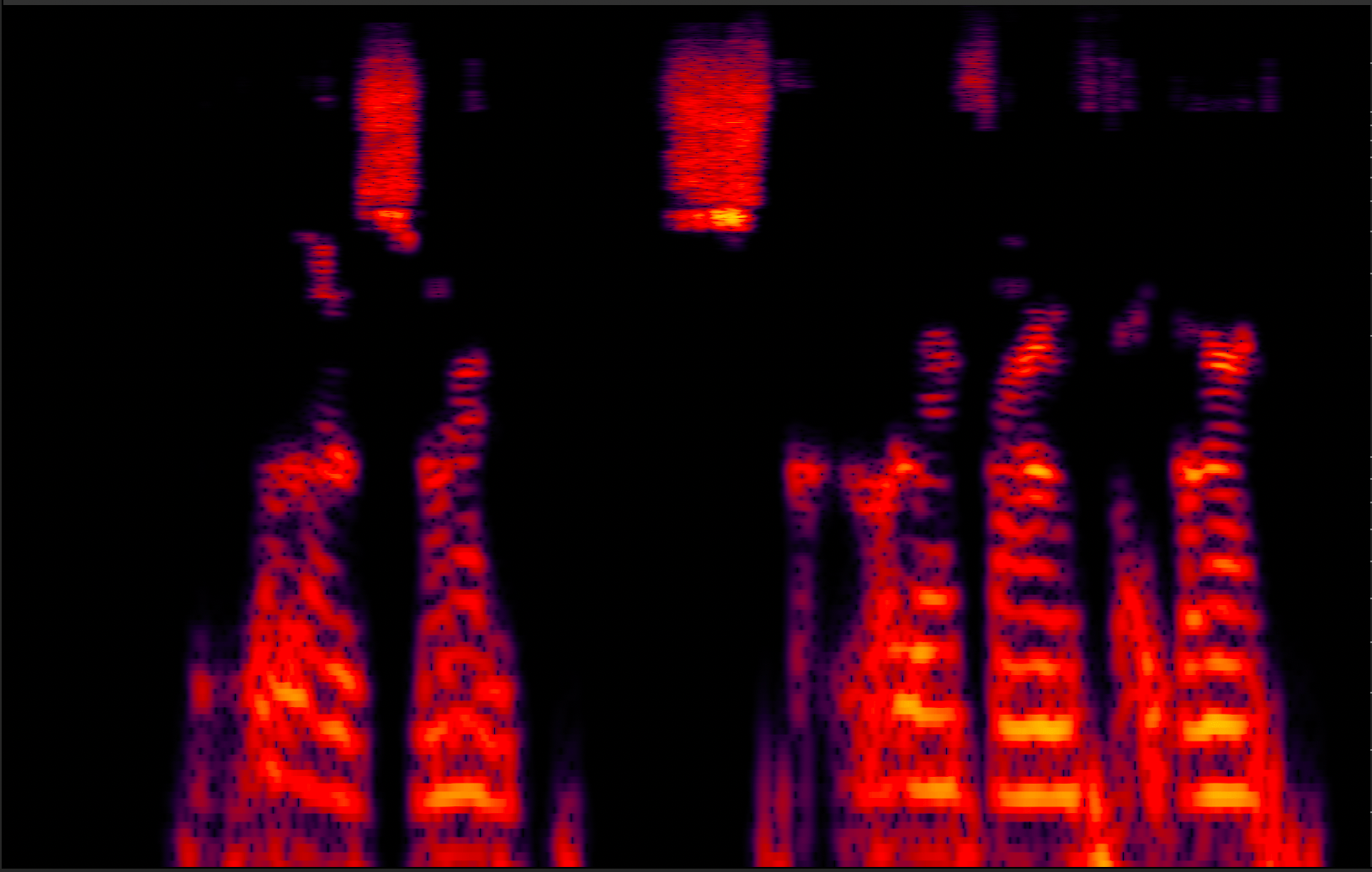} & 
        \includegraphics[align=c,width=0.12\linewidth]{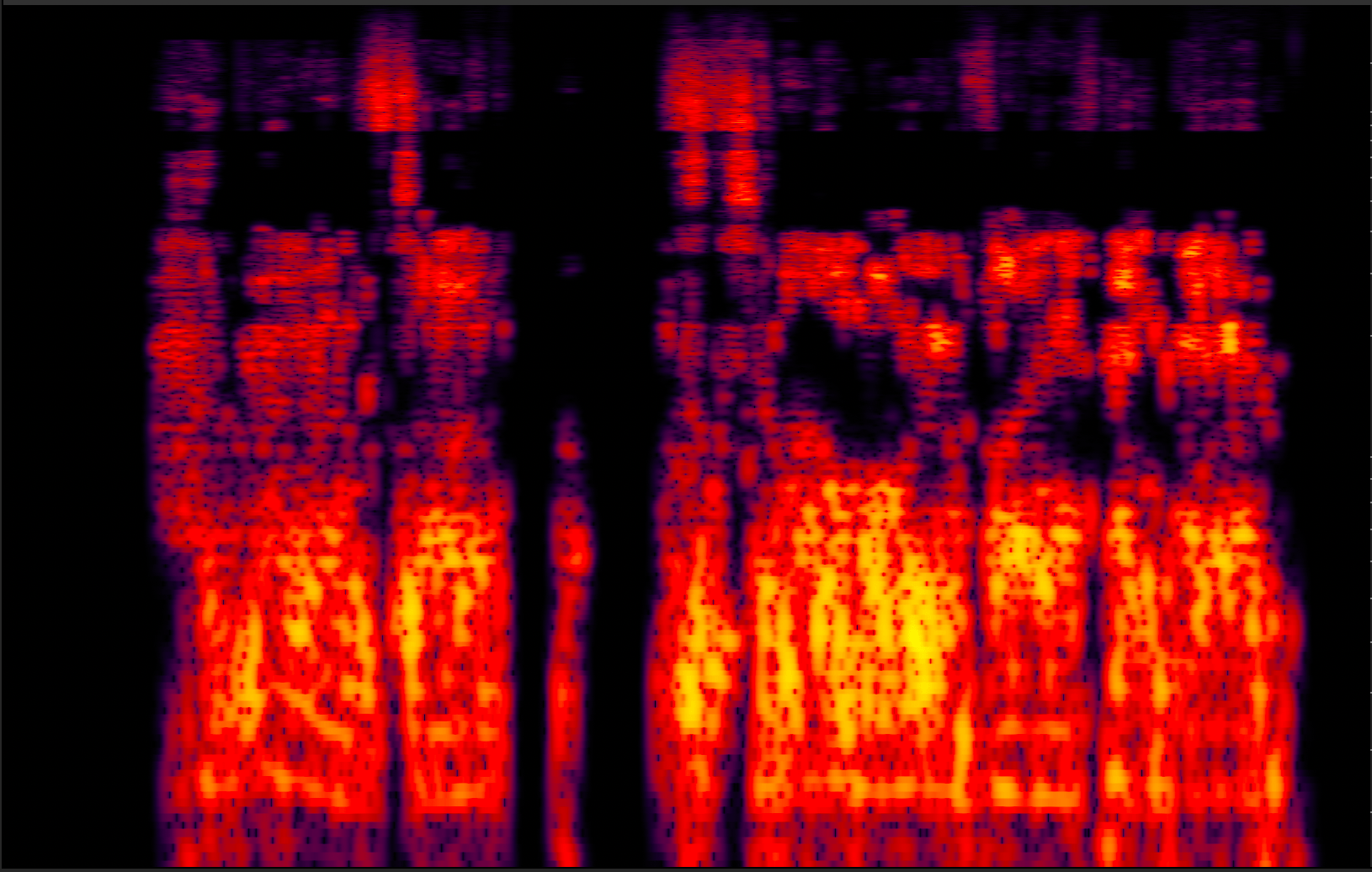} & 
        \includegraphics[align=c,width=0.12\linewidth]{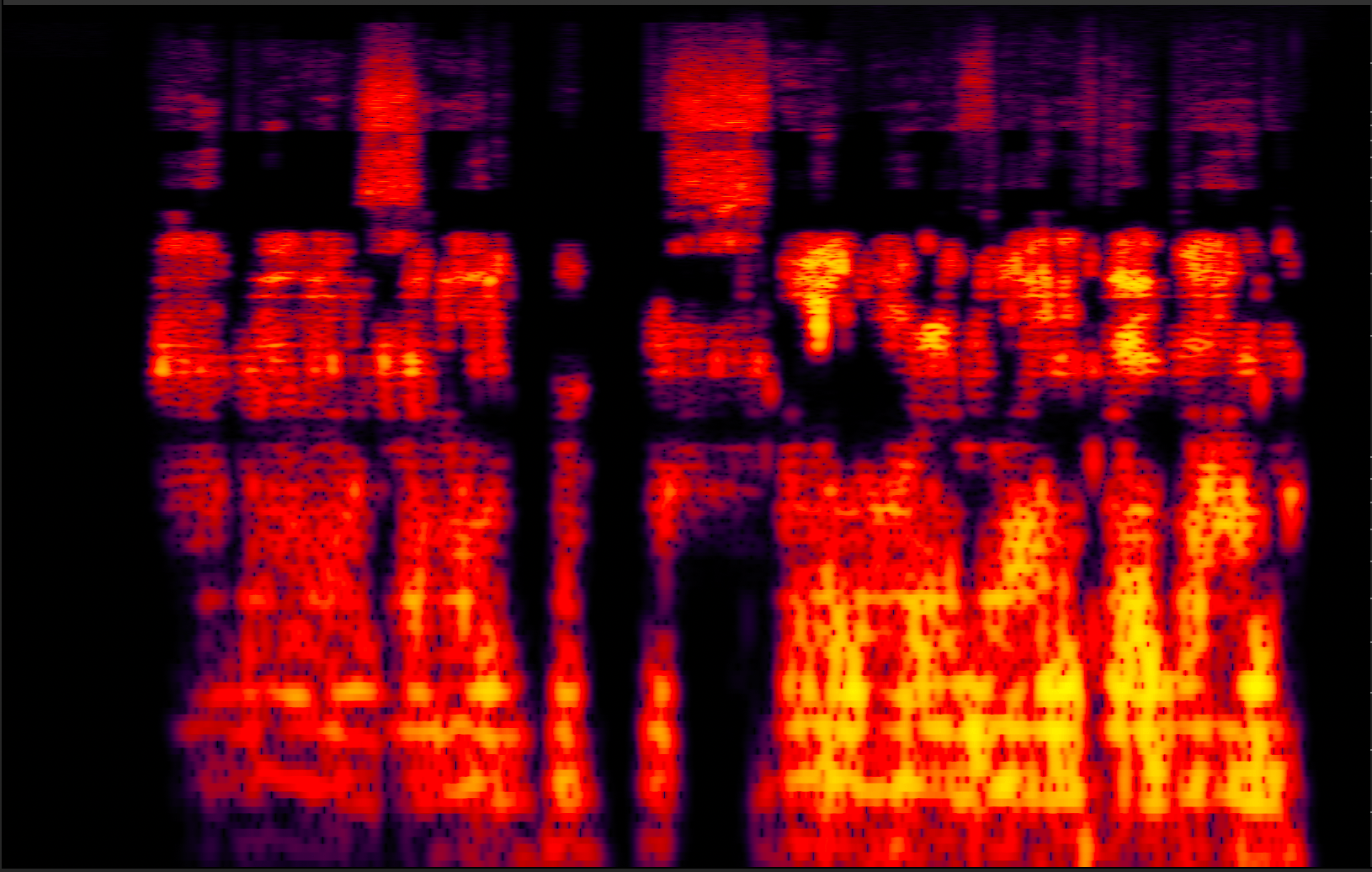} \\ 
        
        \bottomrule
    \end{tabular}
    }
    \caption{Comparison of models using iterative enhancement. Each row represents the number of iterative enhancements applied, where the output of the previous enhancement step is used as input for the next step. The columns show different models being compared.}
    \label{fig:model-comparison}
\end{figure*}

\begin{figure}[b]
  \centering
  \includegraphics[width=0.94\linewidth]{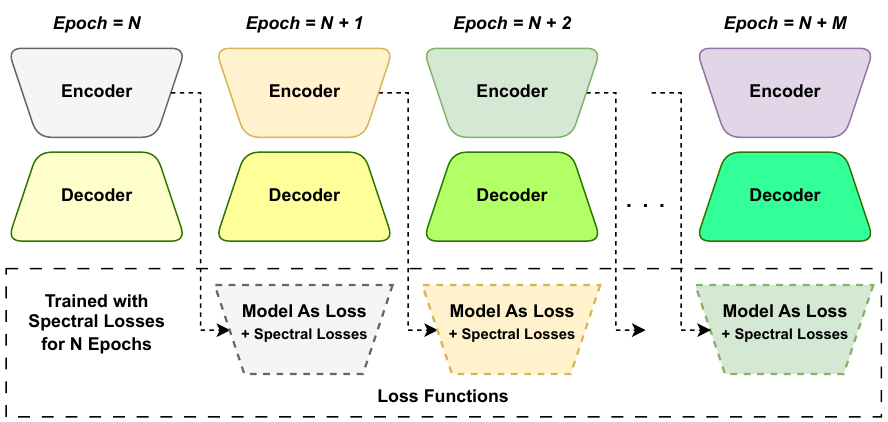}
  \caption{An illustration of the Model as Loss paradigm, showcasing the $\mathcal{L}_{\text{mal}-{\text{dynamic}}}$ variation.}
  \label{fig:mal_figure}
\end{figure}


To address these limitations, we propose a novel training paradigm, \textit{Model as Loss (MAL)}, which uses the encoder of the same model as a loss function to guide the training of the decoder. It involves training a model with conventional loss functions and then using the trained encoder's embeddings as a loss function for the next stage. This approach aligns the loss function with the downstream task, leveraging the encoder's ability to extract task-specific features while ensuring contextual and hierarchical understanding of the signal. Unlike traditional methods that rely on external pre-trained models or handcrafted losses, this paradigm exploits the encoder's specialization in processing noisy speech and its inherent ability to prioritize perceptually relevant signal components.

Our proposed method has several advantages. First, the encoder's feature space is inherently tailored to the enhancement task, capturing both global and local signal features crucial for noise suppression. Second, using the encoder as a loss function enforces a feedback loop that aligns training and inference dynamics, improving generalization to unseen noise types. Third, this approach ensures relevance across all frequency bins by leveraging the encoder's weighting of spectral components based on their contribution to the task, avoiding the pitfalls of uniformly weighted losses. Finally, the self-consistency of the model provides a robust foundation for optimizing the decoder, leading to superior performance and perceptual quality.

In this paper, we present a detailed analysis of the \textit{MAL} paradigm, including its theoretical foundations and practical implementation. Through extensive experiments, we demonstrate that our approach outperforms conventional pre-trained deep feature losses and hand-crafted loss functions on standard speech enhancement benchmarks. The results highlight the potential of using the model itself as a loss function, offering a new perspective on loss design in machine learning.

\section{Methodology}
\label{section:mal}

In the realm of speech enhancement, models typically comprise an encoder and one or more decoders \cite{deepfilternet2}. For simplicity, we will refer to this setup as a single encoder-decoder system moving forward. The encoder's job is to extract relevant features from the noisy signal, which are then used by the decoder to synthesize the enhanced signal.

The usual approach for training encoder-decoder models involves minimizing a loss function $\mathcal{L}$ between the clean reference speech and the model output. For example, an L1 loss between the clean and enhanced signal in spectral domain using Short-time Fourier Transform (STFT) can be represented as:
\begin{equation} \label{eqn_spectral_loss}
	\mathcal{L}_{\text{spectral}} = \Vert \text{STFT}(\mathbf{y}_{\text{clean}}) - \text{STFT}(\mathbf{y}_{\text{enhanced}}) \Vert_1
\end{equation}

where y\textsubscript{clean} is the clean reference speech and y\textsubscript{enhanced} is the model output when the noisy speech is used as input to the model. The objective is to minimize the difference between the model's output and the clean speech, typically measured on a spectrogram or multiple spectrograms of different resolutions.


Once we train a model with some loss $\mathcal{L}$ to convergence, we know that for any given pair (y\textsubscript{enhanced}, y\textsubscript{clean}), $\mathcal{L}$ will be minimal. However, this does not imply that any other loss $\mathcal{L}$\textsubscript{new} will also be minimal. Our search is to find the ideal loss function $\mathcal{L}$\textsubscript{ideal}, such that, once trained to convergence, for any given mathematical function $\mathcal{F}$, we get the minimum loss:
\begin{equation} \label{eqn_f_loss}
	\mathcal{L}_{\text{F}} = \Vert \mathcal{F}(\mathbf{y}_{\text{clean}}) - \mathcal{F}(\mathbf{y}_{\text{enhanced}}) \Vert_1
\end{equation}

Babaev \textit{et al.} \cite{babaev2024finally} propose a \textit{Signal-to-Noise (SNR) Rule}, which suggests that as more noise is added to speech (lowering the SNR), feature representations should move farther apart in the embedding space. Extending this idea, intuitively, once the model is trained to satisfy equation \eqref{eqn_f_loss}, y\textsubscript{enhanced} and y\textsubscript{clean} should be identical points in the embedding space.  Assuming equation \eqref{eqn_f_loss} holds, if we input  y\textsubscript{clean} to the trained model, it should return y\textsubscript{clean}. Moreover, y\textsubscript{enhanced} when used as input should again return y\textsubscript{enhanced}, which is y\textsubscript{clean}. This would ensure the stability of the model in its embedding space. However, this is not the case for most models.
Figure \ref{fig:model-comparison} illustrates the impact of iteratively enhancing the noisy input multiple times with various trained models, which are described later in Section \ref{section:exp-setup}. All models show a clear degradation in speech quality with each iteration, suggesting a loss of fine-grained features.

\begin{table*}[t]
\centering
\caption{NISQA and ScoreQ metrics are presented for both in-domain (In) and out-of-domain (Out) datasets. All proposed MAL-based models demonstrate superior performance, with Ours$_{\text{mal}-\text{dynamic}}$ achieving the highest results across all NISQA metrics.}
\label{tab:nisqa_torchsquim_metrics}
\resizebox{\textwidth}{!}{%
\begin{tabular}{@{}l*{2}{cc}*{2}{cc}*{2}{cc}*{2}{cc}*{2}{cc}*{2}{cc}@{}}
\toprule
\textbf{} & \multicolumn{10}{c}{\textbf{NISQA ($\uparrow$)}} & \multicolumn{4}{c}{\textbf{ScoreQ MOS ($\uparrow$)}} \\
\cmidrule(lr){2-11} \cmidrule(lr){12-15}
\textbf{Model / loss variants} & 
\multicolumn{2}{c}{\textbf{Overall}} & 
\multicolumn{2}{c}{\textbf{Noisiness}} & 
\multicolumn{2}{c}{\textbf{Discontinuity}} & 
\multicolumn{2}{c}{\textbf{Coloration}} & 
\multicolumn{2}{c}{\textbf{Loudness}} & 
\multicolumn{2}{c}{\textbf{Natural}} & 
\multicolumn{2}{c}{\textbf{Synthetic}} \\
& In & Out & In & Out & In & Out & In & Out & In & Out & In & Out & In & Out \\ 
\midrule
Baseline & 
3.50 & 2.93 & 4.06 & 3.82 & 3.87 & 3.45 & 3.45 & 2.97 & 3.90 & 3.50 & 2.91 & 2.50 & 2.10 & 2.01 \\

Baseline$_{\text{10epochs}}$ &
3.54 & 2.99 & 4.10 & 3.85 & 3.92 & 3.53 & 3.45 & 2.99 & 3.89 & 3.53 & 2.93 & 2.51 & 2.09 & 2.00 \\

Baseline$_{\text{wavlm}}$ &
3.56 & 3.02 & 4.06 & 3.85 & 3.94 & 3.55 & 3.47 & 3.03 & 3.89 & 3.57 & 2.93 & 2.52 & 2.07 & 1.99 \\

Baseline$_{\text{wavlm}-\text{fe}}$ &
3.52 & 2.97 & 4.07 & 3.86 & 3.88 & 3.49 & 3.45 & 2.97 & 3.91 & 3.54 & 2.94 & 2.52 & 2.11 & 2.02 \\

\textbf{Ours$_{\text{mal}-\text{frozen-fe}}$} &
3.65 & 3.12 & \textbf{4.14} & 3.93 & 4.01 & 3.60 & 3.57 & 3.10 & 3.98 & 3.62 & \textbf{3.00} & \textbf{2.60} & 2.11 & 2.02 \\

\textbf{Ours$_{\text{mal}-\text{frozen}}$} &
3.66 & 3.13 & 4.09 & 3.90 & 4.06 & 3.64 & 3.58 & 3.15 & 3.99 & \textbf{3.65} & 2.97 & 2.58 & 2.08 & 1.99 \\

\textbf{Ours$_{\text{mal}-\text{dynamic}}$} &
\textbf{3.72} & \textbf{3.17} & \textbf{4.14} & \textbf{3.96} & \textbf{4.10} & \textbf{3.67} & \textbf{3.62} & \textbf{3.16} & \textbf{4.01} & 3.62 & 2.96 & 2.58 & 2.10 & \textbf{2.03} \\

\textbf{Ours$_{\text{wavlm-mal}}$} &
3.65 & 3.13 & 4.13 & 3.95 & 4.02 & 3.61 & 3.57 & 3.11 & 3.97 & 3.61 & 3.00 & \textbf{2.60} & \textbf{2.12} & \textbf{2.03} \\

\midrule
DeepFilterNet2 \cite{deepfilternet2} &
3.46 & 2.82 & 3.97 & 3.66 & 3.87 & 3.47 & 3.43 & 2.88 & 3.86 & 3.48 & 2.90 & 2.49 & 2.08 & 1.98 \\

Ours$_{\text{mal}-{\text{frozen-fe}}}^{\text{ablation}}$ &
3.62 & 3.01 & 4.12 & 3.87 & 3.97 & 3.51 & 3.55 & 3.04 & 3.97 & 3.59 & 2.95 & 2.56 & 2.10 & 2.00 \\

\bottomrule
\end{tabular}%
}
\label{table:nisqa}
\end{table*}

\begin{table*}[t]
\centering
\caption{SIGMOS Metrics are shown across in-domain (In) and out-of-domain (Out) datasets, while Intrusive Metrics are evaluated on the 2024 Urgent Challenge non-blind test set. All proposed MAL-based models outperform the other models.}
\label{tab:sigmos_metrics}
\resizebox{\textwidth}{!}{%
\begin{tabular}{@{}l*{2}{cc}*{2}{cc}*{2}{cc}*{2}{cc}*{2}{cc}*{2}{cc}@{}}
\toprule
\textbf{} & \multicolumn{10}{c}{\textbf{SIGMOS ($\uparrow$)}} & \multicolumn{4}{c}{\textbf{Intrusive Metrics}} \\
\cmidrule(lr){2-11} \cmidrule(lr){12-15}

\textbf{Model / loss variants} & 
\multicolumn{2}{c}{\textbf{Signal}} & 
\multicolumn{2}{c}{\textbf{Overall}} & 
\multicolumn{2}{c}{\textbf{Noise}} & 
\multicolumn{2}{c}{\textbf{Discontinuity}} & 
\multicolumn{2}{c}{\textbf{Coloration}} & 
\textbf{PESQ ($\uparrow$)} & 
\textbf{ESTOI ($\uparrow$)} & 
\textbf{LSD ($\downarrow$)} & 
\textbf{MCD ($\downarrow$)} \\
& In & Out & In & Out & In & Out & In & Out & In & Out & Out & Out & Out & Out \\ 
\midrule

Baseline & 
3.48 & 3.15 & 3.05 & 2.74 & 4.07 & 4.08 & 3.83 & 3.73 & 3.56 & 3.22 & 1.96 & 0.75 & 5.12 & 5.06 \\

Baseline$_{\text{10epochs}}$ &
3.52 & 3.19 & 3.09 & 2.77 & \textbf{4.12} & 4.10 & 3.89 & 3.82 & 3.56 & 3.20 & 2.01 & 0.76 & 5.14 & 4.99 \\

Baseline$_{\text{wavlm}}$ &
3.48 & 3.22 & 3.05 & 2.80 & 4.03 & 4.08 & 3.87 & 3.87 & 3.53 & 3.22 & \textbf{2.03} & 0.76 & 4.99 & 4.91 \\

Baseline$_{\text{wavlm}-\text{fe}}$ &
3.49 & 3.18 & 3.06 & 2.77 & 4.04 & 4.10 & 3.86 & 3.78 & 3.55 & 3.21 & 1.99 & 0.76 & 4.99 & 5.02 \\

\textbf{Ours$_{\text{mal}-\text{frozen-fe}}$} &
\textbf{3.57} & \textbf{3.31} & \textbf{3.14} & 2.88 & 4.10 & 4.12 & 3.92 & 3.93 & \textbf{3.60} & 3.30 & \textbf{2.03} & \textbf{0.77} & 5.02 & 4.88 \\

\textbf{Ours$_{\text{mal}-\text{frozen}}$} &
3.53 & \textbf{3.31} & 3.10 & \textbf{2.89} & 4.04 & 4.11 & 3.91 & \textbf{3.95} & 3.58 & \textbf{3.32} & \textbf{2.03} & \textbf{0.77} & \textbf{4.81} & \textbf{4.87} \\

\textbf{Ours$_{\text{mal}-\text{dynamic}}$} &
3.56 & 3.30 & 3.13 & 2.88 & 4.08 & \textbf{4.13} & 3.91 & 3.91 & 3.59 & \textbf{3.32} & 2.00 & 0.76 & 4.88 & \textbf{4.87} \\

\textbf{Ours$_{\text{wavlm-mal}}$} &
3.55 & 3.30 & 3.11 & 2.87 & 4.08 & 4.11 & \textbf{3.93} & 3.92 & 3.58 & 3.29 & \textbf{2.03} & \textbf{0.77} & 4.98 & 4.92 \\

\bottomrule
\end{tabular}%
}
\label{table:sigmos}
\end{table*}

Since the model being trained is itself a mathematical function, we propose using its encoder as a loss function. After training a model with traditional loss functions, the encoder inherently learns to represent the input data's structure in its feature space. This feature space encodes rich information beyond what traditional loss functions capture. Building on this, we introduce a novel training strategy: first, train the model with conventional losses, and then, using the trained encoder, add a loss term based on the encoder's latent feature space. This loss function compares the encoder's bottleneck embeddings of the enhanced output and the clean signal and can be expressed as:

\begin{equation} \label{eqn_2}
	\mathcal{L}_{\text{mal}} = \Vert Encoder(\mathbf{y}_{\text{clean}}) - Encoder(\mathbf{y}_{\text{enhanced}}) \Vert_1
\end{equation}
where \textit{Encoder} represents the function that returns the bottleneck features, denoted as the \textit{MAL-encoder}. This ensures that the decoder's output aligns with the clean signal not only at the spectral level but also in the encoder's learned feature space. With this formulation, the decoder is guided to preserve the rich feature representation learned by the encoder. As a result, the decoder learns to produce y\textsubscript{enhanced}, which, when fed back into the model, reproduce close to y\textsubscript{enhanced}. This is evident in Figure \ref{fig:model-comparison}, where Ours$_\text{mal}$ models, trained with the \textit{MAL} paradigm, preserve more speech harmonics after 100 iterations.

As shown in Figure \ref{fig:mal_figure}, we train a model with conventional losses for N epochs. Then, for the next M epochs, we add the $\mathcal{L}_{\text{mal}}$ loss to further refine the model. Depending on how we use MAL-encoder, there are three possible $\mathcal{L}_{\text{mal}}$ variations:
\begin{enumerate}
    \item $\mathcal{L}_{\text{mal}-{\text{frozen-fe}}}$: Freeze the trained Encoder (FE) of $N\textsuperscript{th}$ epoch and use it as the \textit{MAL-encoder} to train only the decoder for subsequent epochs.
    \item $\mathcal{L}_{\text{mal}-{\text{frozen}}}$: Use the trained encoder of the $N\textsuperscript{th}$ epoch as \textit{MAL-encoder} for all subsequent epochs and train the full encoder-decoder model.
    \item $\mathcal{L}_{\text{mal}-{\text{dynamic}}}$: Use the trained encoder of the $n\textsuperscript{th}$ epoch as \textit{MAL-encoder} for $(n + 1)\textsuperscript{th}$ epoch for $n \geq N$ and train the full model. Hence, \textit{MAL-encoder} is updated with every epoch.
\end{enumerate}

Within this training paradigm, the encoder (applied to noisy input) functions solely as a feature extractor, while the decoder becomes the primary model for synthesizing the enhanced output. The \textit{MAL-encoder}, acting as a loss function, ensures that the decoder produces an output closely aligned with the features of clean audio. The combination of supervised learning (matching the clean signal) and self-supervised learning (consistency within the encoder's feature space) ensures that the decoder's outputs are both accurate and perceptually meaningful. 

The proposed method introduces a self-consistency feedback loop, where the decoder's output is evaluated not only against the clean signal but also through the encoder's feature space. Such a dual-objective structure reinforces meaningful learning at multiple levels of abstraction, leading to superior performance. This integration of self-supervised and supervised learning balances explicit and implicit learning objectives, making the model more robust in real-world scenarios.


\section{Experimental setup}
\label{section:exp-setup}

We base all our experiments on DeepFilterNet2 proposed by Schröter \textit{et al.} \cite{deepfilternet2}. It has an encoder that extracts relevant features and passes them into a first-stage decoder. The output of this decoder is passed into the deep filtering decoder, which predicts the deep filtering coefficients for each time frame. We train DeepFilterNet2 as the base model from the official GitHub repository \cite{deepfilternet2}, using a 960-point FFT, 480 hop size, Vorbis window and 2-frame lookahead. Training is done on English subset of the 48KHz DNS4 dataset \cite{dubey2022icassp2022deepnoise}, using the same loss function $\mathcal{L}_{\text{df}}$ as in the original article, combining multiple spectral losses. This model serves as the \textbf{Baseline} for our experiments.

The experimental setup consists of five evaluation systems. The first three configurations include finetuning the baseline model with one of the three $\mathcal{L}_{\text{mal}}$ loss variations, mentioned in Section \ref{section:mal}. In each model, the $\mathcal{L}_{\text{mal}}$ loss is added in equal proportion to the original DeepFilterNet2 loss function:\\
i) \textbf{Ours$_{\text{mal}-{\text{frozen-fe}}}$} \hspace{7pt} trained with $\mathcal{L}_{\text{df}}$ +  $\mathcal{L}_{\text{mal}-{\text{frozen-fe}}}$\\
ii) \textbf{Ours$_{\text{mal}-{\text{frozen}}}$} \hspace{11pt} trained with $\mathcal{L}_{\text{df}}$ + $\mathcal{L}_{\text{mal}-{\text{frozen}}}$ \\
iii) \textbf{Ours$_{\text{mal}-{\text{dynamic}}}$} \hspace{3pt} trained with $\mathcal{L}_{\text{df}}$ + $\mathcal{L}_{\text{mal}-{\text{dynamic}}}$

In the next two setups, we introduce a loss function, $\mathcal{L}_{\text{wavlm}}$, which uses the final Conv-layer output from the pre-trained WavLM-Base-Plus (WavLM) \cite{wavlm}, resembling equation \eqref{eqn_2}. We again add this loss function in equal proportion to the original loss function, keeping the encoder either frozen or trainable:\\
iv) \textbf{Baseline$_{\text{wavlm}}$} \hspace{12pt} trained with $\mathcal{L}_{\text{df}}$ + $\mathcal{L}_{\text{wavlm}}$ \\
v) \textbf{Baseline$_{\text{wavlm}-{\text{fe}}}$} \hspace{4pt} trained with $\mathcal{L}_{\text{df}}$ + $\mathcal{L}_{\text{wavlm}-\text{fe}}$

As an ablation experiment to evaluate the effect of $\mathcal{L}_{\text{wavlm}-\text{fe}}$ helps, we combine it with $\mathcal{L}_{\text{mal}-{\text{frozen-fe}}}$ and $\mathcal{L}_{\text{df}}$, adding all three in equal proportions, and denote this model as \textbf{Ours$_{\text{wavlm-mal}}$}.

In all experiments, the \textbf{Baseline} model is finetuned for ten epochs and the best epoch is chosen. For consistency, we also finetune the baseline model for ten epochs without introducing any new losses. This is \textbf{Baseline$_\text{10epochs}$}.

\subsection{Evaluation data}

The evaluation was carried out on a total of 7404 samples drawn from multiple test sets. We divide the test sets into two domains:
\begin{itemize}
    \item \textbf{In-domain test set (3504 samples)}: Since the models are trained on DNS V4 training data, we aggregate the test samples from DNS Challenge V2 \cite{reddy2020interspeech}, V3 \cite{dubey2022icassp2022deepnoise}, and V5 \cite{dubey2023icassp2023deepnoise} covering diverse acoustic scenarios such as mouse clicks, headset noise, speakerphone noise, and emotional speech.
    \item \textbf{Out-of-domain test set (3900 samples)}: We aggregated fully unseen test sets from 2024 and 2025 Urgent Challenges, combining the two nonblind and two blind sets \cite{zhang2024urgent}.
\end{itemize}
This test setup enabled robust performance comparisons across diverse acoustic conditions.

\subsection{Evaluation metrics}

Models are evaluated using SIGMOS, NISQA v2.0, and ScoreQ (no-reference natural and synthetic MOS) metrics \cite{ristea2024icassp2024speechsignal, Mittag_2021, ragano2024scoreq}. SIGMOS scores were computed with all enhanced samples normalized to a peak level of -10 dBFS to account for level dependency. For NISQA, enhanced samples were normalized to have an active speech level of -26 dBFS. Intrusive metrics such as PESQ, ESTOI, log-spectral distance (LSD), and Mel-cepstral distance (MCD) were also calculated. We perform ANOVA analysis and report only the metrics with statistical significance. For intrusive metrics, we use only the 2024 Urgent Challenge nonblind test set with its open-source evaluation pipeline \cite{zhang2024urgent}.


\section{Results}


Tables \ref{table:nisqa} and \ref{table:sigmos} present all metrics, clearly demonstrating that the proposed models with $\mathcal{L}_{\text{mal}}$ losses outperform the others.  Ours$_{\text{mal}-{\text{dynamic}}}$ achieves the best performance across all NISQA metrics, while Ours$_{\text{mal}-{\text{frozen}}}$ leads in all intrusive metrics. The Ours$_{\text{mal}-{\text{frozen-fe}}}$ model outperforms others in SIGMOS Signal and Overall metrics, while performing comparably on the remaining SIGMOS metrics.

Given that Ours$_{\text{mal}-{\text{frozen-fe}}}$ is at par with Ours$_{\text{wavlm-mal}}$, using $\mathcal{L}_{\text{wavlm}-\text{fe}}$ with $\mathcal{L}_{\text{mal}-{\text{frozen-fe}}}$ offers no advantage over $\mathcal{L}_{\text{mal}-{\text{frozen-fe}}}$ alone. Notably, WavLM, with 95.1M parameters, is trained on 94,000 hours of speech data \cite{wavlm}, while DeepFilterNet2, with only 2.31M parameters, is trained on 1,100 hours of speech data \cite{deepfilternet2}. However, all models trained with $\mathcal{L}_{\text{mal}}$ outperform both models trained with $\mathcal{L}_{\text{wavlm}}$ loss variants.

Typically, finetuning leads to overfitting on in-domain data, resulting in degraded performance on out-of-domain samples \cite{Li_2022}. This is particularly a risk when the \textit{MAL-encoder} is optimized specifically for in-domain data. However, the results demonstrate that Ours$_{\text{mal}}$ models not only avoid overfitting but also significantly outperform all other models.

\subsection{Ablation Experiments}

A key factor is the quality of the \textit{MAL-encoder} for $\mathcal{L}_{\text{mal}-{\text{frozen}}}$ or $\mathcal{L}_{\text{mal}-{\text{frozen-fe}}}$, as the effectiveness of the loss function depends on how well the encoder extracts features for enhancement. To answer this, we used the publicly available pre-trained DeepFilterNet2 \cite{deepfilternet2}, which performs slightly worse than our trained Baseline. We finetuned the pre-trained model with $\mathcal{L}_{\text{mal}-{\text{frozen-fe}}}$ as previously described, resulting in Ours$_{\text{mal}-{\text{frozen-fe}}}^{\text{ablation}}$. Although $\mathcal{L}_{\text{mal}-{\text{frozen-fe}}}$ significantly improved performance, it still performed worse than when applied to the superior Baseline model (see Table \ref{table:nisqa}). The better the encoder, the more effective $\mathcal{L}_{\text{mal}-{\text{frozen-fe}}}$ becomes. 

\begin{figure}[t]
  \centering
  \includegraphics[width=0.905\linewidth]{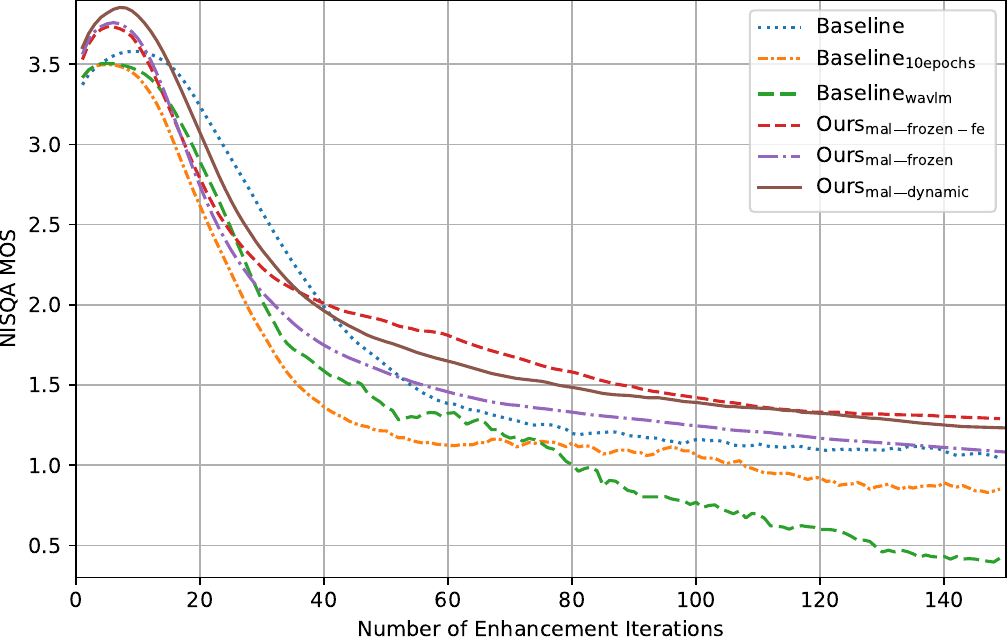}
  \caption{NISQA MOS vs number of enhancement iterations}
  \label{fig:itr_test}
\end{figure}

We also trained with $\mathcal{L}_{\text{mal}-{\text{dynamic}}}$ per batch instead of per epoch, resulting in slightly worse metrics (e.g., LSD 5.03 vs. 4.88), likely due to loss instability from frequent updates.\\
\textbf{Self-consistency Experiment}: We take the first 200 samples of the 2025 Urgent Challenge nonblind test set and iteratively enhance them 150 times with every model. Figure \ref{fig:itr_test} shows how the average NISQA MOS initially improves as noise is removed but later declines as speech quality degrades. Models trained with $\mathcal{L}_{\text{mal}-{\text{frozen-fe}}}$ or $\mathcal{L}_{\text{mal}-{\text{dynamic}}}$ better preserve speech, aligning with the self-consistency criterion, which $\mathcal{L}_{\text{mal}-{\text{frozen}}}$ lacks. In the 1\textsuperscript{st} iteration, all \textit{MAL} models achieve a high MOS, then reach a higher peak before converging to a higher MOS.

This aligns with Figure \ref{fig:model-comparison}, where \textit{MAL} models preserve more speech harmonics than non-\textit{MAL} models. Notably, the Baseline model preserves slightly more speech than Baseline$_{\text{10epoch}}$ or Baseline$_{\text{wavlm}}$, despite having lower metrics (see Tables \ref{tab:nisqa_torchsquim_metrics} and \ref{tab:sigmos_metrics}). Although the exact reason is unclear, we suspect that the latter two models are more aggressive than the Baseline in removing noise.

\section{Conclusion}

In this paper, we propose \textit{Model as Loss (MAL)}, a novel training paradigm that leverages the encoder of an encoder-decoder model as a loss function to guide optimization. By aligning the loss with the model's task-specific feature space, \textit{MAL} overcomes the limitations of traditional handcrafted and pre-trained deep feature losses. This approach offers key advantages such as task-specific feature extraction, self-consistency, and enhanced contextual understanding of input signals. Additionally, \textit{MAL} reduces dependency on deep-feature losses derived from pre-trained models while achieving comparable or superior performance. This is particularly essential in domains such as medical imaging \cite{gondara2016medical} or specialized signal analysis \cite{sabharwal2024enhancing}, where pre-trained models are scarce. Experiments demonstrate that \textit{MAL} improves both perceptual quality and task-specific performance in speech enhancement.

Although our evaluation focused on speech enhancement, \textit{MAL}'s domain-agnostic design makes it applicable to tasks such as acoustic echo cancellation \cite{zhang2022neural, braun2022task}, image denoising or super-resolution \cite{tian2020deep}, and medical image analysis \cite{li2021review}. Ongoing research explores its broader potential. We hope that this work inspires further innovation in loss functions and training methodologies, advancing machine learning applications.

\section{Acknowledgements}
The authors thank Paul Kendrik, Tijana Stojkovic, and Andy Pearce for their valuable feedback and insights. We also thank Sai Dhawal Phaye for discussions during the early stages of \textit{MAL}, and Kanav Sabharwal for his feedback on the writing.
\newpage

\bibliographystyle{IEEEtran}
\bibliography{mybib}

\balance

\end{document}